\documentclass[preprint,aps,prd,showpacs,nofootinbib,superscriptaddress,floatfix,dvipdfmx]{revtex4-1}

\usepackage{amsmath}
\usepackage{amssymb}
\usepackage{graphicx}
\usepackage{ascmac}
\usepackage{url}

\usepackage{hyperref}
\hypersetup{%
 setpagesize=false,%
 bookmarks=true,%
 bookmarksdepth=tocdepth,%
 bookmarksnumbered=true,%
 colorlinks=true,%
 allcolors=blue,
 pdftitle={},%
 pdfsubject={},%
 pdfauthor={},%
 pdfkeywords={}}
 
\def\VEV#1{
\langle #1\rangle
}
\begin{document}
\title{Tension between neutrino masses and gauge coupling unification in natural grand unified theories}

\author{Nobuhiro Maekawa}
\email[]{maekawa@eken.phys.nagoya-u.ac.jp}
\affiliation{Department of Physics,
Nagoya University, Nagoya 464-8602, Japan}
\affiliation{Kobayashi-Maskawa Institute for the Origin of Particles and the
Universe, Nagoya University, Nagoya 464-8602, Japan}
\affiliation{KEK Theory Center, Tsukuba 305-0801, Japan}

\author{Taiju Tanii}
\email[]{tanii.t@eken.phys.nagoya-u.ac.jp}
\affiliation{Department of Physics,
Nagoya University, Nagoya 464-8602, Japan}

\date{\today}

\begin{abstract}
    \noindent
The natural grand unified theories solve various problems of the supersymmetric grand unified theory and give realistic quark and lepton mass matrices under the natural assumption that all terms allowed by the symmetry are introduced with $O(1)$ coefficients. However, because of the natural assumption, it is difficult to achieve the gauge coupling unification without tuning, 
while keeping neutrino masses at realistic values. In this paper, we try to avoid this tension
between the neutrino masses and the gauge coupling unification, by introducing suppression factors
for several terms. These suppression factors can be understood by approximate symmetries in some
of the solutions. 
We show that one of the most important results in the natural GUT scenario, that the nucleon decay mediated by superheavy gauge fields is enhanced due to smaller unification scale while the nucleon decay mediated by superheavy colored Higgs is suppressed, can change in those models proposed in this paper.

\end{abstract}

\maketitle
\section{Introduction}
The grand unified theory (GUT) \cite{Georgi:1974sy} is the most promising model beyond the standard model (SM).
The GUT not only unifies three of the four forces in nature except gravity, but also unifies matter such as quarks and leptons. Furthermore, the GUT already has been supported from experiments for each unification.
With respect to force unification, it has been shown that by introducing supersymmetry (SUSY) \cite{SUSY}, the three running gauge couplings in the SM coincide on the grand unified scale $\Lambda_G\sim 2\times 10^{16}$ GeV \cite{threegaugeunification}. As for the unification of matter, the various hierarchies of quark and lepton masses and mixings measured by various experiments \cite{Workman:2022ynf} can be reasonably explained by the unification of matter in the $SU(5)$ GUT, in which one generation quarks and leptons can be unified into $\bf\bar 5$ and $\bf 10$ of $SU(5)$. In $SU(5)$ unification, the single assumption, that $\bf 10$ fields induce stronger hierarchies in Yukawa couplings than $\bf\bar 5$ field, explains not only
various mass hierarchies of quarks and leptons - the up-type quarks have the strongest mass hierarchy, the neutrinos have the weakest mass hierarchy
\footnote{
Here, we assume the normal hierarchy for neutrino mass spectrum, and we do not take account
of the lightest neutrino mass which has not been observed yet.
}, 
and the down-type quarks and the charged leptons have the intermediate  mass hierarchies, but also, simultaneously,  that the quark mixings are smaller than the lepton mixings. This is a non-trivial result,
and hence this can be understood as the experimental support for the matter unification in $SU(5)$ unification.
Unfortunately, SUSY GUT has two main problems. The first is that the unification of matter also unifies the mass matrices of quarks and leptons, which are inconsistent with the observed values \cite{Langacker:1980js}. The second problem is that the mass of the SM Higgs partner must be sufficiently large compared to the SM Higgs mass to obtain sufficiently stable proton \cite{Randall:1995sh}, which is difficult to achieve without fine tuning.

The natural GUTs \cite{NaturalSO10,Maekawa:2002mx,NaturalE6} solve those SUSY GUT problems under the natural assumption that all terms allowed by the symmetry are introduced with $O(1)$ factors.  As a result, we have a natural GUT that becomes the SM at low energy. Unfortunately, the problem is that in order to achieve unification of the gauge coupling constants, many $O(1)$ coefficients in the range of 0.5 to 2 must be artificially choosed, 
which in a sense is a fine tuning. As will be discussed in detail in this paper, this issue is related with the measured neutrino masses. That is, under the natural assumptions in the natural GUT, there is a tension between the unification of gauge coupling constants and the measured neutrino masses.

In this paper, we discuss how to resolve this tension. In particular, we consider the possibility of solving this problem by considering cases in which small suppression factors, rather than O(1) factors, are applied to several particular terms. After building the models, we discuss the origin of those small suppression factors, such as approximate symmetries. 
Moreover, we discuss the predictions of nucleon decay in those models. 
Interestingly, the predictions of the natural GUT, that nucleon decay via dimension 6 operators is enhanced due to smaller unification scale and nucleon decay via dimension 5 operators is suppressed, can change in those models. 

After this introduction, we review the natural GUT in section II. In section III, we discuss the solutions to solve this tension and approximate symmetries to make this solution natural. We also build explicit natural GUT models and discuss the nucleon decay in these GUT models. Section IV is for discussions and summary.

%%%%%%%%%%%%%%%%%%%%%%%%%%%%%%%%
\section{Natural GUT and its problems}
%%%%%%%%%%%%%%%%%%%%%%%%%%%%%%%%

In this section, we review the $SO(10)$ natural GUT \cite{NaturalSO10,Maekawa:2002mx} and its problem, which also appears in the 
$E_6$ natural GUT \cite{NaturalE6}. %is more general for the natural GUT %. Let us explain them 
%with the natural $SO(10)$ GUT. %We explain them using $SO(10)$ GUT, but it is a discussion for the general Natural GUT.
One of the most important features in the natural GUT is that all terms allowed by the symmetry are introduced with $O(1)$ coefficients.  Because of this feature, once we fix the symmetry of the model, the definition of the model can be done except $O(1)$ coefficients. Under the symmetry $SO(10)\times U(1)_A\times Z_2$, typical quantum numbers of 
the field contents are given in Table \ref{SO10}.
\begin{table}[t]
  \begin{center}
    \begin{tabular}{|c||c|c|c|}
      \hline
      $SO(10)$ & negatively charged fields & positively charged fields & matter fields \\ \hline \hline
      {\bf 45}  & $A(a=-1,-)$ & $A'(a'=3,-)$ & \\ \hline
      {\bf 16} & $C(c=-4,+)$ & $C'(c'=3,-)$  & $\Psi_i(\psi_1=\frac{9}{2}, \psi_2=\frac{7}{2}, \psi_3=\frac{3}{2}, +)$ \\ \hline
      ${\bf \overline{16}}$ & $\bar C(\bar c=-1,+)$ & $\bar C'(\bar c'=7,-)$ & \\ \hline
      {\bf 10} & $H(h=-3,+)$  & $H'(h'=4,-)$ & $T(t=\frac{5}{2},+)$ \\ \hline
      1 &  $\left.\begin{array}{c}\Theta(\theta=-1,+), \\ Z(z=-2,-), \bar Z(\bar z=-2,-) \end{array}\right.$ & $S(s=5,+)$ &
      \\ \hline
    \end{tabular}
    \caption{Field contents of natural $SO(10)$ GUT with $U(1)_A$ charges. $\pm$ shows $Z_2$ parity. The half integer $U(1)_A$ charges play the same role as R-parity.}
    \label{SO10}
  \end{center}
\end{table}
In this paper, we use large characters for fields or operators and small characters for their $U(1)_A$ charges.
The $U(1)_A$ \cite{U(1)} has gauge anomalies which are cancelled by the Green-Schwarz mechanism \cite{Green-Schwartz}
\footnote{ 
Strictly, the fields in Table \ref{SO10} alone may not satisfy the conditions for the anomaly cancellation, but the arbitrariness of the normalization of $U(1)_A$ and the introduction of a few new $SO(10)$ singlet fields can satisfy those conditions.
}, and the Fayet-Iliopoulos (FI) term \cite{Fayet:1974jb} $\xi^2\int d^2\theta V_A$ is assumed, where $V_A$ is a vector multiplet of the $U(1)_A$. 
It is surprising that the various problems in SUSY GUT scenarios, including the doublet-triplet splitting problem,  can be solved in this model with the above natural feature. 
Unfortunately, there is a tension between the neutrino masses and the unification of the gauge couplings. 
Let us explain them in details in the review of the natural GUT below.

\subsection{Anomalous $U(1)_A$ gauge symmetry}
First, for simplicity, we consider a simpler model in which we have only three matter fields
$\Psi_i$, and two negatively charged fields $H$, and $\Theta$ in Table \ref{SO10}. 
The superpotential invariant under $U(1)_A$ is given as
\begin{equation}
\label{SO10Y}
W_Y=c_{ij}\sum_{i,j=1,2,3}\left(\frac{\Theta}{\Lambda}\right)^{\psi_i+\psi_j+h}\Psi_i\Psi_jH,
\end{equation}
where $\Lambda$ and $c_{ij}$ are  the cutoff of the model and $O(1)$ coefficients. 
If we assume that only $\Theta$ has a non-vanishing vacuum expectation value (VEV), which  is determined by the $D$-flatness condition of $U(1)_A$ as $\langle\Theta\rangle=\xi\equiv\lambda\Lambda$,
the interaction terms in the above superpotential becomes the hierarchical Yukawa interactions as
\begin{equation}
W_Y=c_{ij}\lambda^{\psi_i+\psi_j+h}\Psi_i\Psi_jH,
\end{equation}
when $\lambda<1$. In this paper, we take $\lambda\sim 0.22$, which is approximately the  Cabibbo angle. 
The realization of the hierarchical structures of Yukawa couplings from higher dimensional effective interactions by developing the VEV of some fields, which breaks a (flavor) symmetry,  is often called the Froggatt-Nielsen mechanism \cite{Froggatt-Nielsen}. % and is often used.
The important point is that Yukawa hierarchies can be reproduced under the natural assumption that all terms allowed by the symmetry are introduced, including higher dimensional terms. 
However, it is quite rare to adopt this natural assumption even in the GUT Higgs sector in which the GUT group is spontaneously broken into the SM gauge group, mainly
because it is difficult to control infinite number of higher dimensional terms.
Within the same theory, it is not reasonable for the Yukawa sector to adopt this natural assumption and the Higgs sector not.
%It is reasonable that even in the GUT Higgs sector, the natural assumption is adopted.
%Since even the various couplings except the $O(1)$ coefficients can be determined by the symmetry, the superpotential for the Higgs sector in GUTs is also determined by the symmetry under this natural assumption.
The natural GUTs are the theories in which this natural assumption is adopted even in the GUT Higgs sector as well as in the Yukawa sector.

Note that the Higgs mass term $\lambda^{2h}H^2$ is forbidden when $h<0$ because of the holomorphic feature of the superpotential, that is called the SUSY zero mechanism, or the holomorphic zero mechanism. 
The SUSY zero mechanism plays important roles in controling  the infinite
number of higher dimensional terms and in solving the doublet-triplet (DT) splitting problem. 
We will explain them in the next subsection.

\subsection{Higgs sector in natural GUT}
In this subsection, we will briefly review the GUT Higgs sector in the natural GUT, which breaks
$SO(10)$ into $G_{SM}\equiv SU(3)_C\times SU(2)_L\times U(1)_Y$ and solves the DT splitting problem under the natural assumption.

One of the most important assumptions is that all positively $U(1)_A$ charged fields have 
vanishing VEVs. This assumption not only allows the SUSY zero mechanism to work, but also to control an infinite number of higher dimensional terms. Under this assumption, it is easy to show that the F-flatness conditions for negatively charged fields are automatically satisfied. The F-flatness conditions of positively charged fields determine  the VEVs of negatively charged fields. Thus, ignoring the D-flatness  conditions, if the number of positively charged fields equals the number of negatively charged fields, the VEVs of all negatively charged fields can be determined in principle. 
In order to break $SO(10)$ into the SM gauge group $G_{SM}$, an adjoint Higgs
${\bf 45}_A$ and one pair of spinor ${\bf 16}_C$ and anti-spinor ${\bf\bar {16}}_{\bar C}$ are
minimally required. In addition, the standard model Higgs ${\bf 10_H}$ is needed. These fields must have negative $U(1)_A$ charges because these fields have non-vanishing VEVs. Moreover, the same number of positively charged fields are introduced as in Table \ref{SO10} to fix the VEVs of these negatively charged fields. It is non-trivial that simply introducing the smallest number of fields necessary in $SO(10)$ GUT as in Table \ref{SO10} can solve various problems, including the DT splitting problem. 
Note that the terms which include two or more positively charged fields has no effects in fixing these VEVs of negatively charged fields under the assumption. Therefore, only the terms which include one positively charged field are important to fix the VEVs. 
The superpotential for fixings the VEVs are
\begin{equation}
W=W_{H'}+W_{A'}+W_{S}+W_{C'}+W_{\bar C'},
\end{equation}
where $W_X$ denotes the terms linear in the $X$ field. Each $W_X$ includes finite number of terms because of the SUSY zero mechanism. Note that only finite number of terms are important to fix the VEVs although the infinite number of higher dimensional terms are introduced.

Now we discuss how to determine the VEVs by $W_X$. First, we consider $W_{A'}$, which is given as
\begin{equation}
W_{A^\prime}=\lambda^{a^\prime+a}A^\prime A+\lambda^{a^\prime+3a}(
(A^\prime A)_{\bf 1}(A^2)_{\bf 1}
+(A^\prime A)_{\bf 54}(A^2)_{\bf 54}),
\end{equation}
where the subscripts {\bf 1} and {\bf 54} denote the representation 
of the composite
operators under the $SO(10)$ gauge symmetry. Unless otherwise noted, the O(1) coefficients are omitted and we take $\Lambda=1$ in this paper.  The F-flatness condition of $A'$ fixes the VEV of
$A$. One of the 6 vacua
\footnote{
Without loss of generality, the VEV is written as $\langle A\rangle=i\tau_2\times {\rm diag}(x_1,x_2,x_3,x_4,x_5)$. The vacua are
classified by the number of 0 because the F-flatness condition of $A'$
gives the solution $x_i=0, v$. Therefore, the number of the vacua becomes 6.
}
becomes the Dimopoulos-Wilczek (DW) form \cite{DW} as 
$\langle A\rangle=i\tau_2\times {\rm diag}(v,v,v,0,0)$ which breaks 
$SO(10)$ into $SU(3)_C\times SU(2)_L\times SU(2)_R\times U(1)_{B-L}$.  Note that the $v$ is determined
by the $U(1)_A$ charge of $A$ as $v\sim \lambda^{-a}$.  
This VEV of $A$ plays an important role in solving the DT splitting problem.
Actually, through
\begin{equation}
W_{H'}=\lambda^{h+a+h'}H'AH
\end{equation}
the triplet Higgses become massive while the doublet Higgses remain massless. 
One pair of doublet Higgses becomes massive through the mass term $\lambda^{2h'}H'^2$.
(Note that to determine the mass spectrum, the terms which include two positively charged fields must be considered. ) Then only one pair of doublet Higgses becomes massless, and therefore, the DT splitting problem can be solved. The effective colored Higgs mass for nucleon decay becomes $\lambda^{2h}$ which is larger than the cutoff scale because
$h<0$. 

The VEVs of $C$ and $\bar C$, which is important to break $SU(3)_C\times SU(2)_L\times SU(2)_R\times U(1)_{B-L}$ into $G_{SM}$, are induced by the F-flatness condition
of $S$ from the superpotential
\begin{equation}
\label{S}
W_S=\lambda^{s}S\left(1+\lambda^{c+\bar c}\bar CC+\sum_k\lambda^{2ka}A^{2k}\right).
\end{equation}
Since $\lambda^{2ka}\langle A^{2k}\rangle\sim 1$, basically the last term in Eq. (\ref{S}) does not change the following result.
% when $O(1)$ coefficients are omitted. 
The F-flatness condition
of $S$ gives $\langle \bar CC\rangle\sim \lambda^{-(c+\bar c)}$, and thus the D-flatness
condition of $SO(10)$ leads to $|\langle C\rangle|=|\langle \bar C\rangle|\sim \lambda^{-(c+\bar c)/2}$.
Note that the VEVs of $C$ and $\bar C$ are determined by their charges again.
The $F$-flatness conditions of $C^\prime$ and 
$\bar C^\prime$ realize the alignment of the VEVs $\langle{C}\rangle$, $\langle{\bar C}\rangle$, and $\langle A\rangle$, 
and impart masses on the pseudo Nambu-Goldstone fields.\footnote{
Without $W_{C'}$ and $W_{\bar C'}$, the superpotential for fixing $\langle A\rangle$ becomes independent of the superpotential for fixing the VEVs $\langle{C}\rangle$, $\langle{\bar C}\rangle$. It means that accidental global symmetry appears and as a result, pseudo Nambu-Goldstone fields appear by breaking the global symmetry.
} 
This mechanism
proposed by Barr and Raby \cite{BarrRaby} is naturally embedded in the natural GUT.
$W_{C'}$ and $W_{\bar C'}$ are given as
\begin{eqnarray}
W_{C^\prime}&=&
       \bar C(\lambda^{\bar c +c^\prime+a}A
       +\lambda^{\bar c +c^\prime+\bar z}\bar Z)C^\prime, \\
W_{\bar C^\prime}&=&
       \bar C^\prime(\lambda^{\bar c^\prime +c+a} A
       +\lambda^{\bar c^\prime +c+z}Z)C.
\end{eqnarray}
Since the VEV of $A$ is proportional to the $B-L$ generator $Q_{B-L}$, only one of 
four component fields $({\bf 3},{\bf 2},{\bf 1})_{1/3}$, 
$({\bf \bar 3},{\bf 1},{\bf 2})_{-1/3}$, 
$({\bf 1},{\bf 2},{\bf 1})_{-1}$ and $({\bf 1},{\bf 1},{\bf 2})_{1}$ 
under
$SU(3)_C\times SU(2)_L\times SU(2)_R\times U(1)_{B-L}$, which are obtained by decomposition of $\bf 16$ of $SO(10)$, has non-vanishing VEV.  When the component
$({\bf 1},{\bf 1},{\bf 2})_{1}$ has non-vanishing VEV, $G_{SM}$ can be obtained. 

\subsection{Mass spectrum of superheavy particles}
Since all the interactions are determined by the symmetry, mass spectrum of superheavy particles are also fixed except $O(1)$ coefficients in the natural GUTs. The mass spectrum are important in calculating the renormalization group equations (RGEs) for the gauge couplings.
Note that we have to consider also the terms which include two positively charged fields in order to
examine the mass spectrum.

The spinor ${\bf 16}$, the vector ${\bf 10}$ and the adjoint ${\bf 45}$ of $SO(10)$
are decomposed under $SO(10)\supset SU(5) \supset SU(3)_C\times SU(2)_L\times U(1)_Y$ as
\begin{eqnarray}
{\bf 16}&\rightarrow &
\underbrace{[Q+U^c+E^c]}_{\bf 10}+\underbrace{[D^c+L]}_{\bf \bar 5}
+\underbrace{N^c}_{\bf 1},\\
{\bf 10}&\rightarrow &
\underbrace{[D^c+L]}_{\bf \bar 5}+\underbrace{[\bar D^c+\bar L]}_{\bf 5},\\
{\bf 45}&\rightarrow &
\underbrace{[G+W+X+\bar X+N^c]}_{\bf 24}
+\underbrace{[Q+U^c+E^c]}_{\bf 10}
+\underbrace{[\bar Q+\bar U^c+\bar E^c]}_{\bf \overline{10}}
+\underbrace{N^c}_{\bf 1},
\end{eqnarray}
where the quantum numbers of $G_{SM}$ are explicitly written as $Q({\bf 3,2})_{\frac{1}{6}}$,
$U^c({\bf \bar 3,1})_{-\frac{2}{3}}$, $D^c({\bf \bar 3,1})_{\frac{1}{3}}$,
$L({\bf 1,2})_{-\frac{1}{2}}$, $E^c({\bf 1,1})_1,N^c({\bf 1,1})_0$,
$X({\bf 3,2})_{-\frac{5}{6}}$,
$G({\bf 8,1})_0$, and $W({\bf 1,3})_0$.

First, let us consider the mass spectrum of $\bf 5$ and $\bf\bar 5$ of $SU(5)$. %The mass matrices for $\bf 5$ and $\bf\bar 5$ fields can be written as
The mass matrices $M_I$ ($I=D^c(H_T),L(H_D)$) can be written as
\begin{equation}
M_I=\bordermatrix{
       &I_H & I_{H^\prime} & I_{C} &I_{C^\prime} \cr
\bar I_H & 0 & \lambda^{h+h^\prime}\alpha_I & 0 & 0 \cr
\bar I_{H^\prime} & \lambda^{h+h^\prime}\alpha_I & \lambda^{2h^\prime} 
& 0 & \lambda^{h'+c'+\frac{1}{2}(c-\bar c)} \cr
\bar I_{\bar C}& 0 & \lambda^{h'+\frac{3}{2}\bar c-\frac{1}{2}c} & 0 
& \lambda^{\bar c+c^\prime}\beta_I \cr
\bar I_{\bar C^\prime} &  \lambda^{h+\bar c'-\frac{1}{2}(c-\bar c)} & \lambda^{h'+\bar c'-\frac{1}{2}(c-\bar c)}
 & \lambda^{c+\bar c^\prime}\beta_I 
  & \lambda^{c^\prime+\bar c^\prime} \cr}.
\end{equation}
where $\alpha_{L(H_D)}$ is vanishing and $\alpha_{D^c(H_T)}\sim O(1)$, while
$\beta_I=\frac{3}{2}((B-L)_I-1)$; that is, $\beta_L=-3$ and $\beta_{D^c}=-1$. Only one pair of
doublet Higgs becomes massless, which comes from 
\begin{equation}
{\bf 5}_H,\quad {\bf\bar 5}_H+\lambda^{h-c+\frac{1}{2}(\bar c-c)}{\bf\bar 5}_C.
\end{equation}

Next, we consider the mass matrices for $\bf 10$ of $SU(5)$, which 
are given by
%The mass matrices are written as $4\times 4$ matrices,
\begin{equation}
M_I=\bordermatrix{
        &I_A & I_{A^\prime} &I_{ C}&I_{C^\prime} \cr
\bar I_A &0& \lambda^{a^\prime+a} \alpha_I & 0  & 
\lambda^{c^\prime-\frac{1}{2}(c-\bar c)+a} \cr
\bar I_{A^\prime} &\lambda^{a+a^\prime} \alpha_I & \lambda^{2a^\prime} & 0 & 
\lambda^{c^\prime-\frac{1}{2}(c-\bar c)+a^\prime} \cr
\bar I_{\bar C}&0 & 0 & 0 & \lambda^{\bar c+c^\prime}\beta_I \cr
\bar I_{\bar C^\prime} &\lambda^{\bar c^\prime+\frac{1}{2}(c-\bar c)+a} &
\lambda^{\bar c^\prime+\frac{1}{2}(c-\bar c)+a^\prime}&
\lambda^{c+\bar c^\prime}\beta_I & \lambda^{c^\prime+\bar c^\prime}\cr}.
\label{mass10}
\end{equation}
Here,  $\alpha_Q$ and $\alpha_{U^c}$ are vanishing because
these are Nambu-Goldstone modes, but $\alpha_{E^c}\sim O(1)$. Also $\beta_Q=-1$, 
$\beta_{U^c}=-2$ and $\beta_{E^c}=0$. 
Thus the each $4\times 4$ matrix has one 
vanishing eigenvalue.
%, which corresponds to the Nambu-Goldstone mode eaten by the Higgs mechanism. 
The mass spectrum of the remaining three
modes is ($\lambda^{c+\bar c^\prime}$, $\lambda^{c^\prime+\bar c}$,
$\lambda^{2a^\prime}$) for $Q$ and $U^c$, and
($\lambda^{a+a^\prime}$, 
$\lambda^{a+a^\prime}$,
$\lambda^{c^\prime+\bar c^\prime}$) or 
($\lambda^{\bar c^\prime+\frac{1}{2}(c-\bar c)+a}$, 
$\lambda^{c^\prime-\frac{1}{2}(c-\bar c)+a}$,
$\lambda^{2a^\prime}$) for $E^c$. 

%The mass matrices are written as $4\times 4$ matrices,
%\begin{equation}
%M_I=\bordermatrix{
%        &I_A & I_{A^\prime} &I_{ C}&I_{C^\prime} \cr
%\bar I_A &0& \lambda^{a^\prime+a} \alpha_I & 0  & 
%\frac{\lambda^{\bar c+c^\prime+a}}{\sqrt{2}}\VEV{\bar C} \cr
%\bar I_{A^\prime} &\lambda^{a+a^\prime} \alpha_I & \lambda^{2a^\prime} & 0 & 
%\frac{\lambda^{\bar c+c^\prime+a^\prime}}{\sqrt{2}}\VEV{\bar C} \cr
%\bar I_{\bar C}&0 & 0 & 0 & \lambda^{\bar c+c^\prime+a}\beta_Iv \cr
%\bar I_{\bar C^\prime} &\frac{\lambda^{c+\bar c^\prime+a}}{\sqrt{2}}\VEV{C} &
%\frac{\lambda^{c+\bar c^\prime+a^\prime}}{\sqrt{2}}\VEV{C} &
%\lambda^{c+\bar c^\prime+a}\beta_Iv & \lambda^{c^\prime+\bar c^\prime}\cr}.
%\label{mass10}
%\end{equation}
%where $\alpha_I$ vanishes for $I=Q$ and $U^c$ because
%these are Nambu-Goldstone modes, but $\alpha_{E^c}\neq 0$.
%On the other hand, $\beta_I=\frac{3}{2}((B-L)_I-1)$; that is, $\beta_Q=-1$, 
%$\beta_{U^c}=-2$ and $\beta_{E^c}=0$. 
%Thus for each $I$, the $4\times 4$ matrix has one 
%vanishing eigenvalue, which corresponds to the Nambu-Goldstone mode eaten
%by the Higgs mechanism. The mass spectrum of the remaining three
%modes is ($\lambda^{c+\bar c^\prime+a}v$, $\lambda^{c^\prime+\bar c+a}v$,
%$\lambda^{2a^\prime}$) for the color-triplet modes $Q$ and $U^c$, and
%($\lambda^{a+a^\prime}$, 
%$\lambda^{a+a^\prime}$,
%$\lambda^{c^\prime+\bar c^\prime}$) or 
%($\lambda^{c+\bar c^\prime+a}\VEV{C}$, 
%$\lambda^{c^\prime+\bar c+a}\VEV{\bar C}$,
%$\lambda^{2a^\prime}$) for the color-singlet modes $E^c$. 

Finally, we consider the mass spectrum for $\bf 24$ of $SU(5)$.
%The adjoint fields $A$ and $A^\prime$ contain
%two $G$, two $W$ and two pairs of $X$ and $\bar X$, 
%whose 
The mass matrices $M_I(I=G,W,X)$ are given by
\begin{equation}
M_I=\bordermatrix{
            &    I_A       &        I_{A'}           \cr
\bar I_A    &     0        & \lambda^{a+a'}\alpha_I  \cr
\bar I_{A'} & \lambda^{a+a'}\alpha_I & \lambda^{2a'}  \cr}.
\end{equation}
The mass spectrum for $G$ and $W$ are $(\lambda^{a^\prime+a},\lambda^{a^\prime+a})$, while for $X$ it becomes
$(0, \lambda^{2h'})$. The massless mode of $X$ 
%Since $\alpha_X=0$, one pair of $X$ is massless, which 
is eaten
by the Higgs mechanism.
%, while the other pair has a rather light mass of
%$\lambda^{2a^\prime}$.

\subsection{Gauge coupling unification}
Since all symmetry breaking scales and all the mass spectrums of superheavy particles
%except those from matter sector 
are
fixed by anomalous $U(1)_A$ charges, we can calculate the running gauge couplings and discuss the gauge coupling unification. 
%\footnote{We have one pair of $\bf 5$ and $\bf\bar 5$ superheavy fields from matter sector, whose masses also determined by their charges as discussed in the next subsection. Those fields do not affect the discussion in this subsection because the mass spectrumrespects $SU(5)$ symmetry. }. 
In this paper, we study the running gauge couplings obtained by one-loop RGEs. Note that superheavy particles from the matter sector which is discussed in the next subsection are complete multiplets of $SU(5)$, and therefore, they do not affect the conditions for unification of the gauge coupling constants. 

In the natural $SO(10)$ GUT, $SO(10)$ is broken by the VEV $\VEV{A}\equiv \Lambda_A\sim \lambda^{-a}$ into $SU(3)_C\times SU(2)_L\times SU(2)_R\times U(1)_{B-L}$, which
is broken by the VEVs $|\VEV{C}|=|\VEV{\bar C}|\equiv\Lambda_C\sim \lambda^{-(c+\bar c)/2}$ into $G_{SM}$. 

Now let us discuss the conditions of the gauge coupling unification 
\begin{equation}
\alpha_3(\Lambda_A)=\alpha_2(\Lambda_A)=
\frac{3}{5}\alpha_Y(\Lambda_A)\equiv\alpha_1(\Lambda_A),
\end{equation}
where 
$\alpha_1^{-1}(\mu>\Lambda_C)\equiv 
\frac{3}{5}\alpha_R^{-1}(\mu>\Lambda_C)
+\frac{2}{5}\alpha_{B-L}^{-1}(\mu>\Lambda_C)$ with the renormalization scale $\mu$. 
Here $\alpha_X\equiv\frac{g_X^2}{4\pi}$ and 
$g_X  (X=3,2,R,B-L,Y)$ are the gauge couplings of 
$SU(3)_C$, $SU(2)_L$, $SU(2)_R$, $U(1)_{B-L}$ and $U(1)_Y$, 
respectively. Since the model has the left-right symmetry above $\Lambda_C$, we expect $g_2=g_R$ at $\mu>\Lambda_C$.

The gauge couplings at the scale $\Lambda_A$ are obtained by one-loop RGEs as
\begin{eqnarray}
\alpha_1^{-1}(\Lambda_A)&=&\alpha_1^{-1}(M_{SB})
+\frac{1}{2\pi}\left(b_1\ln \left(\frac{M_{SB}}{\Lambda_A}\right)
+\Sigma_i \Delta b_{1i}\ln \left(\frac{m_i}{\Lambda_A}\right)
-\frac{12}{5}\ln \left(\frac{\Lambda_C}{\Lambda_A}\right)\right), 
\label{alpha1}\\
\alpha_2^{-1}(\Lambda_A)&=&\alpha_2^{-1}(M_{SB})
+\frac{1}{2\pi}\left(b_2\ln \left(\frac{M_{SB}}{\Lambda_A}\right)
+\Sigma_i \Delta b_{2i}\ln \left(\frac{m_i}{\Lambda_A}\right)\right), \\
\alpha_3^{-1}(\Lambda_A)&=&\alpha_3^{-1}(M_{SB})
+\frac{1}{2\pi}\left(b_3\ln \left(\frac{M_{SB}}{\Lambda_A}\right)
+\Sigma_i \Delta b_{3i}\ln \left(\frac{m_i}{\Lambda_U}\right)\right), 
\end{eqnarray}
where $M_{SB}$ is a SUSY breaking scale. 
Here, $(b_1,b_2,b_3)=(33/5,1,-3)$ represent the 
renormalization group coefficients
for the minimal SUSY standard model(MSSM)
and $\Delta b_{ai}(a=1,2,3)$ denote the corrections to these coefficients 
arising from the massive fields with mass $m_i$, which can be read from the Table 
\begin{center}
\begin{tabular}{|c|c|c|c|c|c|c|c|c|}
\hline
$I$      & $Q+\bar Q$ & $U^c+\bar U^c$ & $E^c+\bar E^c$ & $D^c+\bar D^c$ 
         & $L+\bar L$ & $G $& $W$ & $X+\bar X$ \\
\hline
$\Delta b_{1I}$ & $\frac{1}{5}$& $\frac{8}{5}$& $\frac{6}{5}$& $\frac{2}{5}$
         & $\frac{3}{5}$ & 0 &  0 & 5 \\
\hline
$\Delta b_{2I}$ & 3 & 0 & 0 & 0 & 1 & 0 & 2 & 3 \\
\hline
$\Delta b_{3I}$ & 2 & 1 & 0 & 1 & 0 & 3 & 0 & 2 \\ 
\hline
\end{tabular}
\end{center}
The last term in Eq. (\ref{alpha1}) is caused by the breaking 
$SU(2)_R\times U(1)_{B-L}\rightarrow U(1)_Y$ due to the VEV $\VEV{C}$.
The gauge couplings at the SUSY breaking scale $M_{SB}$
can be obtained by the success of the gauge coupling unification in the 
MSSM as
\begin{eqnarray}
\alpha_1^{-1}(M_{SB})&=&\alpha_G^{-1}(\Lambda_G)
+\frac{1}{2\pi}\left(b_1\ln \left(\frac{\Lambda_G}{M_{SB}}\right)\right),\\
\alpha_2^{-1}(M_{SB})&=&\alpha_G^{-1}(\Lambda_G)
+\frac{1}{2\pi}\left(b_2\ln \left(\frac{\Lambda_G}{M_{SB}}\right)\right), \\
\alpha_3^{-1}(M_{SB})&=&\alpha_G^{-1}(\Lambda_G)
+\frac{1}{2\pi}\left(b_3\ln \left(\frac{\Lambda_G}{M_{SB}}\right)\right),
\end{eqnarray}
where $\alpha_G^{-1}(\Lambda_G)\sim 25$ and 
$\Lambda_G\sim 2\times 10^{16}$ GeV.
The above conditions for unification are rewritten as
\begin{eqnarray}
&&\left(\frac{\Lambda_A}{\Lambda_G}\right)^{14}
\left(\frac{\Lambda_C}{\Lambda_A}\right)^6
\left(\frac{\det \bar M_L}{\det \bar M_{D^c}}\right)
\left(\frac{\det \bar M_Q}{\det \bar M_{U}}\right)^4
\left(\frac{\det \bar M_Q}{\det \bar M_{E^c}}\right)^3
\left(\frac{\det \bar M_W}{\det \bar M_{X}}\right)^5  \\ \nonumber
&=& \Lambda_A^{-\bar r_{D^c}+\bar r_L-4\bar r_{U^c}-3\bar r_{E^c}+7\bar r_Q
-5\bar r_X+5\bar r_W}, \\
&&\left(\frac{\Lambda_A}{\Lambda_G}\right)^{16}
\left(\frac{\Lambda_C}{\Lambda_A}\right)^4
\left(\frac{\det \bar M_{D^c}}{\det \bar M_{L}}\right)
\left(\frac{\det \bar M_Q}{\det \bar M_{U}}\right)
\left(\frac{\det \bar M_Q}{\det \bar M_{E^c}}\right)^2
\left(\frac{\det \bar M_G}{\det \bar M_{X}}\right)^5  \\ \nonumber
&=&\Lambda_A^{-\bar r_{L}+\bar r_{D^c}-\bar r_{U^c}-2\bar r_{E^c}+3\bar r_Q
-5\bar r_X+5\bar r_G}, \\
&&\left(\frac{\Lambda_A}{\Lambda_G}\right)^{4}
\left(\frac{\det \bar M_{D^c}}{\det \bar M_{L}}\right)
\left(\frac{\det \bar M_U}{\det \bar M_{Q}}\right)
\left(\frac{\det \bar M_G}{\det \bar M_{W}}\right)^2
\left(\frac{\det \bar M_G}{\det \bar M_{X}}\right) \\ \nonumber
&=&\Lambda_A^{-\bar r_{L}+\bar r_{D^c}-\bar r_{Q}+\bar r_{U}-2\bar r_W
-\bar r_X+3\bar r_G}, 
\end{eqnarray}
where $\bar M_I$ are the reduced mass matrices where massless modes are omitted from the original mass matrices %which have no massless mode 
and $\bar r_I$ are rank of the reduced mass matrices. 
In our scenario, the symmetry breaking scales 
%the unification scale 
$\Lambda_A\sim \lambda^{-a}$, 
%the symmetry breaking scale 
$\Lambda_C\sim \lambda^{-\frac{1}{2}(c+\bar c)}$, 
and the  determinants of 
the reduced mass matrices are determined by the anomalous $U(1)_A$ charges;
\begin{eqnarray}
\det \bar M_{Q}&\sim &\det \bar M_{U^c}\sim 
\lambda^{2a'+c+\bar c+c'+\bar c'}, \label{detQ} \\
\det \bar M_{E^c}&\sim &\lambda^{2a+2a'+c'+\bar c'}, \\
\det M_{D^c}&\sim &\lambda^{2h+2h'+c+\bar c+c'+\bar c'}, \\
\det \bar M_{L}&\sim &\lambda^{2h'+c+\bar c+c'+\bar c'}, \\
\det M_{G}&\sim &\det M_W\sim \lambda^{2a+2a'}, \\
\det \bar M_{X}&\sim &\lambda^{2a'}. \label{detX}
\end{eqnarray}
The unification conditions $\alpha_1(\Lambda_A)=\alpha_2(\Lambda_A)$,
$\alpha_1(\Lambda_A)=\alpha_3(\Lambda_A)$ and
$\alpha_2(\Lambda_A)=\alpha_3(\Lambda_A)$ lead to
$\Lambda\sim \lambda^{\frac{h}{7}}\Lambda_G$,
$\Lambda\sim \lambda^{-\frac{h}{8}}\Lambda_G$ and 
$\Lambda\sim \lambda^{-\frac{h}{2}}\Lambda_G$, respectively.
So the unification conditions become 
\begin{eqnarray}
h&\sim& 0,\\
%, and then the cutoff scale must be taken as $
\Lambda &\sim &\Lambda_G.
\end{eqnarray}
Surprisingly, the above unification conditions do not depend on the anomalous $U(1)_A$ charges except $h$. 
%Note that these relations are independent of the anomalous $U(1)_A$ charges except that of the doublet Higgs. 
This can be shown to be a general result in the GUT with the anomalous $U(1)_A$ 
\cite{Maekawa:2002mx}. 
It is important that the cutoff scale in the natural GUT is taken to be around the usual GUT scale. It means that the true GUT scale $\Lambda_A\equiv \langle A\rangle\sim \lambda^{-a}\Lambda$ becomes smaller than $\Lambda_G$. Therefore, the nucleon decay via superheavy gauge field exchange is enhanced and it may be seen in near future experiments.

%implies that this result can be applied
%to rather general cases. 
Unfortunately, in the natural GUT model in Table \ref{SO10}, we take $h=-3$ not $h=0$.
Of course, to forbid the explicit SM Higgs mass term $H^2$, $h$ must be negative.
But only because of that, we can take larger $h$, for example, $h=-1$.
We take $h=-3$ in order to obtain realistic neutrino masses. In other word, if we take
$h\sim 0$, the neutrino masses become too small.
We will explain them in the next subsection.

%On the other hand, we should not take this relation $h\sim 0$ seriously 
%because we have an ambiguity 
%of order one coefficients and use only one loop renormalization group 
%equations. However, in order to catch the tendency, the above analysis
%is fairly useful. 

\subsection{Matter sector in natural $SO(10)$ GUT}
In this subsection, we will briefly review how to obtain realistic quark and lepton masses and mixings in the natural $SO(10)$ GUT. Especially, neutrino masses will be explained in details because we will introduce a tension between the neutrino masses and gauge coupling unification condition later.

If the Yukawa interactions have been obtained only from the superpotential in Eq. (\ref{SO10Y}), the model would be unrealistic because of the unrealistic $SO(10)$ GUT relations
\begin{equation}
Y_u=Y_d=Y_e=Y_{\nu_D}
\end{equation}
for the Yukawa couplings. The easiest way to avoid this unrealistic $SO(10)$ GUT relations is to introduce $\bf 10$ of $SO(10)$ as a matter field in addition to three $\bf 16$ as in Table \ref{SO10}. The model has four $\bf\bar 5$ and one $\bf 5$ of $SU(5)$ since
$\bf 16$ and $\bf 10$ of $SO(10)$ are decomposed under $SU(5)$ as
$\bf 16=10+\bar 5+1$ and $\bf 10=5+\bar 5$.  One of four $\bf\bar 5$s becomes superheavy with a $\bf 5$ field through the interactions
\begin{equation}
W=\lambda^{\psi_i+t+c}\Psi_iTC+\lambda^{2t}T^2.
\end{equation}
Main modes of three massless $\bf\bar 5$ fields become $(\bf\bar 5_{\Psi_1},\bar 5_T,\bar 5_{\Psi_2})$, and the Yukawa matrices are obtained as
\begin{equation}
Y_u=\left(
\begin{array}{ccc}
\lambda^6 & \lambda^5 & \lambda^3 \\
\lambda^5 & \lambda^4 & \lambda^2 \\
\lambda^3 & \lambda^2 & 1    
\end{array}
\right),\quad
Y_d\sim Y_e^T\sim Y_{\nu_D}^T\sim \lambda^2\left(
\begin{array}{ccc}
\lambda^4 & \lambda^{3.5} & \lambda^3 \\
\lambda^3 & \lambda^{2.5} & \lambda^2 \\
\lambda^1 & \lambda^{0.5} & 1    
\end{array}
\right).
\label{quark}
\end{equation}
Note that the higher dimensional interactions, 
$\lambda^{\psi_i+\psi_j+c+\bar c+h}\Psi_i\Psi_j\bar CCH$ and 
$\lambda^{\psi_i+\psi_j+2La+h}\Psi_i\Psi_jA^{2L}H$, give the same order contributions to
these Yukawa couplings as 
$\lambda^{\psi_i+\psi_j+h}\Psi_i\Psi_jH$
after developing the VEVs, $\langle \bar CC\rangle\sim\lambda^{-(c+\bar c)}$ and $\langle A\rangle\sim\lambda^{-a}$.  Because of this feature, the $SU(5)$ GUT relation
$Y_d=Y_e^T$ can naturally be avoided in the natural GUT. 
Thus, 
we can obtain
the Cabibbo-Kobayashi-Maskawa (CKM) \cite{Kobayashi:1973fv} matrix as
\begin{equation}
U_{\rm CKM}=
\left(
\begin{array}{ccc}
1 & \lambda &  \lambda^3 \\
\lambda & 1 & \lambda^2 \\
\lambda^3 & \lambda^2 & 1
\end{array}
\right),
\label{CKM}
\end{equation}
which is consistent with the experimental value if we choose 
$\lambda\sim 0.22$\footnote{Here, it is also important that the massless Higgs doublet come from 
$\bar 5_H+\lambda^{h-c+\frac{1}{2}(\bar c-c)}\bar 5_C$.}.

The right-handed neutrino masses are obtained from the interactions
\begin{equation}
\lambda^{\psi_i+\psi_j+2\bar c}\Psi_i\Psi_j\bar C\bar C
\end{equation}
as
\begin{equation}
M_R=\lambda^{\psi_i+\psi_j+2\bar c}\VEV{\bar C}^2
=\lambda^{2\psi_3+\bar c-c}\left(
\begin{array}{ccc}
\lambda^6 & \lambda^5 & \lambda^3 \\
\lambda^5 & \lambda^4 & \lambda^2 \\
\lambda^3   & \lambda^2         & 1
\end{array}
\right).
\end{equation}
Thus, the neutrino mass matrix is written as
\begin{equation}
M_\nu=M_{\nu_D}M_R^{-1}M_{\nu_D}^T=\lambda^{4-2\psi_3+c-\bar c}\left(
\begin{array}{ccc}
\lambda^2          & \lambda^{1.5}  & \lambda \\
\lambda^{1.5} & \lambda & \lambda^{0.5} \\
\lambda            & \lambda^{0.5}  & 1 
\end{array}
\right)\VEV{H_u}^2.
\label{neutrino}
\end{equation}
The Maki-Nakagawa-Sakata (MNS) matrix \cite{Maki:1962mu} is obtained from the $Y_e$ in eq. (\ref{quark}) and $M_\nu$ in eq.(\ref{neutrino}) as
\begin{equation}
 U_{MNS}= \left(
\begin{array}{ccc}
1         & \lambda^{0.5}  & \lambda \\
\lambda^{0.5} & 1 & \lambda^{0.5} \\
\lambda            & \lambda^{0.5}  & 1 
\end{array}
\right).
\label{MNS}
\end{equation}
%where $\eta$ is a renormalization factor. 
To obtain the observed neutrino masses \cite{Workman:2022ynf}, 
\begin{equation}
\lambda^{4-2\psi_3+c-\bar c}\frac{\VEV{H_u}^2}{\Lambda}\sim m_{\nu_\tau}\sim 0.05{\rm eV}
\end{equation}
is required.  When $\Lambda\sim \Lambda_G\sim 2\times 10^{16}{\rm GeV}$, this condition is
rewritten as 
\begin{equation}
    h+c-\bar c\sim -6, 
\end{equation}
which is satisfied with the natural GUT in Table \ref{SO10}. Here we use $h+2\psi_3=0$, which
is required to obtain $O(1)$ top Yukawa coupling. It is difficult to obtain larger $h$ by smaller
$c$ and/or larger $\bar c$, because several conditions are required to obtain realistic natural
GUT models as 
\begin{enumerate}
  \item $2\psi_3+h=0$: to obtain $O(1)$ top Yukawa coupling, i.e., to obtain the term $\lambda^0\Psi_3\Psi_3H$.
  \item $t-\psi_3-\frac{1}{2}(c-\bar c)=\frac{5}{2}$:  To obtain the MNS matrix as in eq. (\ref{MNS}), which is consistent with the observations.
  \item $\psi_3+t+c\geq 0$: To allow the term $\Psi_3TC$ which make ${\bar 5}_{\Psi_3}$ superheavy.
  \item $\psi_3+\psi_1+2{\bar c}\geq 0$: To allow the term $\Psi_3\Psi_1{\bar C}^2$ which makes the right-handed neutrino mass matrix's rank three.
  \item $c+{\bar c}+a'+a<0$: To forbid the term ${\bar C}A'AC$ which destabilizes the DW type VEV.
  \item $c+{\bar c}\geq -6$: To obtain realistic quark masses and mixings. 
\end{enumerate}
Especially conditions 2,3, and 5 are critical for the difficulty to obtain larger $h$.
This is the tension between the neutrino masses and the gauge coupling unification in the natural
GUT scenario.
The above conditions are considered to build explicit natural GUT models with suppression factors
in the next section. Note that we have assumed that $\psi_1=\psi_3+3$ and $\psi_2=\psi_3+2$ to obtain realistic quark and lepton masses and mixings.

\section{Solutions for tension between neutrino masses and gauge coupling unification}
In this section, we examine several possibilities to avoid the tension between the neutrino masses and the gauge coupling unification.
Since this tension is strongly dependent on the basic assumption that all terms allowed by the symmetry are introduced with $O(1)$ coefficients, 
we explore the possibilities in which some of the terms have much smaller coefficients than 1.
When we introduce the terms with small coefficients, we require that the VEV relations for the GUT singlet operator $O$ with the $U(1)_A$ charge $o$ as
\begin{equation}
\langle O\rangle\sim \lambda^{-o}
\label{VEV}
\end{equation}  
do not change because they play critical roles in the natural GUT scenario.

After finding the set of terms with small coefficients that avoids this tension, we discuss the reason for the small coefficients, for example, 
because of an approximate symmetry etc.

Furthermore, we build concrete natural GUT models which can avoid the tension between the neutrino masses and the gauge coupling unification.
And we discuss the nucleon decay in those models.
%We construct models that can simultaneously explain gauge coupling unification and neutrino mass by tuning the coefficients of superpotential.

%%%%%%%%%%%%%%%%%%%%%%%%%%%%%%%%%%%%%%%%%%
\subsection{Suppression factor for terms for right-handed neutrino masses}
\label{nu coefficient}
One of the easiest way to avoid the tension is to introduce small coefficients for the terms which give the right-handed neutrino masses as
%We consider the coefficient of the following term, which is the origin of the mass of the right-handed neutrino.
\begin{equation}
\varepsilon_{\nu}\lambda^{\psi_i+\psi_j+2 \bar{c}} \Psi_i \Psi_j \bar{C} \bar{C},
\end{equation}
where we omit the $O(1)$ coefficients. Since the right-handed neutrino masses become smaller, the (left-handed) neutrino 
masses become larger. The heaviest neutrino mass can be given as %is the tau neutrino, the mass is determined  by
\begin{equation}
m_{\nu_{\tau}}=\frac{1}{\varepsilon_\nu}\lambda^{4+h+c-{\bar c}}\frac{\langle H_u\rangle^2}{\Lambda},
\label{neutrino3}
\end{equation} 
which must be the observed value $m_{\nu_{\tau}}\sim 0.05$ eV. 
Note that this suppression factor does not change the VEV relation in eq. (\ref{VEV}), and the mass
spectrum of superheavy particles except the right-handed neutrinos. This means that the beta functions do not change,
and therefore, the gauge coupling unification conditions remain unchanged as $h\sim 0$ and $\Lambda=\Lambda_G$. 
Since $h=0$ allows the Higgs
mass term $H^2$ which spoils the doublet-triplet splitting, we take $h=-1$. An concrete natural GUT model with $h=-1$
is given in Table \ref{assignment_nu2}. Note that the half integer $U(1)_A$ charges for matter fields play the same
role as the R-parity, and all requirements listed in the end of the previous section are satisfied in this model.
%\begin{table}[htbp]
%  \centering
%  \begin{equation}
%  \begin{array}{c|c|c|c} 
%  & \text { Negative charged fields } & \text { Positive charged fields } & \text { matter fields }\\
%  \hline \mathbf{4 5} & A(a=-1,-) & A^{\prime}\left(a^{\prime}=3,-\right) & \\
%  \mathbf{1 6} & C(c=-2,+) & C^{\prime}\left(c^{\prime}=3,-\right) &\Psi_i(\psi_1=\frac{7}{2},\psi_2=\frac{5}{2},\psi_3=\frac{1}{2},+)\\
%  \overline{\mathbf{1 6}} & \bar{C}(\bar{c}=-1,+) & \bar{C}^{\prime}\left(\bar{c}^{\prime}=4,-\right) & \\
%  \mathbf{1 0} & H(h=-1,+) & H^{\prime}\left(h^{\prime}=2,-\right) &T(t=\frac{5}{2},+)\\
%  \mathbf{1} & Z(z=-2,-), \bar{Z}(\bar{z}=-2,-) & S(s=5,+) & \\
%  \end{array}\nonumber
%  \end{equation}
%  \caption{In the case of $(t,c,{\bar c})=\left(\frac{5}{2},-2,-1\right)$ in Model 1}
%  \label{assignment_nu1}
%  \end{table}
%  
\begin{table}[t]
  \begin{center}
    \begin{tabular}{|c||c|c|c|}
      \hline
      $SO(10)$ & negatively charged fields & positively charged fields & matter fields \\ \hline \hline
      {\bf 45}  & $A(a=-1,-)$ & $A'(a'=3,-)$ & \\ \hline
      {\bf 16} & $C(c=-3,+)$ & $C'(c'=4,-)$  & $\Psi_i(\psi_1=\frac{7}{2}, \psi_2=\frac{5}{2}, \psi_3=\frac{1}{2}, +)$ \\ \hline
      ${\bf \overline{16}}$ & $\bar C(\bar c=-2,+)$ & $\bar C'(\bar c'=5,-)$ & \\ \hline
      {\bf 10} & $H(h=-1,+)$  & $H'(h'=2,-)$ & $T(t=\frac{5}{2},+)$ \\ \hline
      1 &  $Z(z=-2,-), \bar Z(\bar z=-2,-)$ & $S(s=5,+)$ &
      \\ \hline
    \end{tabular}
    \caption{In the case of $(t,c,{\bar c})=\left(\frac{5}{2},-3,-2\right)$ in Model 1.}
    \label{assignment_nu2}
  \end{center}
\end{table}
  
From the eq.(\ref{neutrino3}), the heaviest neutrino mass is given by 
$m_{\nu_{\tau}}=\frac{1}{\varepsilon_\nu}\lambda^{2}\frac{\langle H_u\rangle^2}{\Lambda}\sim 0.05$ eV,
which determines the suppression factor $\varepsilon_\nu\sim 10^{-3}$.

The effective colored Higgs mass for the nucleon decay becomes $m_{H_C}^{eff}\sim \lambda^{2h}\Lambda\sim 10^{18}$ GeV,
which results in sufficient suppression of nucleon decay via colored Higgs medidation.
On the other hand, since the nucleon decay via gauge boson
mediation is enhanced, it can be seen in near future experiments 
as in the usual natural GUT scenario. 

Unfortunately, we have not found an approximate symmetry to understand this suppression factor. We need other reasoning for this suppression factor.

% to satisfy the gauge coupling unification condition, $\varepsilon_{\nu}\sim 0.01$ gives the observed 
%neutrino masses. 
%takes correct size at $h\sim-1$. 
%We are assuming that the cutoff scale is $\Lambda\sim2\times10^{18}{\rm GeV}$. Then, if we take $\langle H_u\rangle=175{\rm GeV},\lambda=0.22$ and $h=-1$, we can obtain $m_{\nu_\tau}\sim\frac{1}{\varepsilon_\nu}13.5\times10^{-6}{\rm eV}$. To be consistent with experimental data, we should choose $\varepsilon_\nu\sim2.7\times10^{-4}$ and then $m_{\nu_\tau}\sim5\times10^{-2}{\rm eV}$.

%%%%%%%%%%%%%%%%%%%%%%%%%%%%%%%%%%%%%%%%%%
\subsection{Suppression factors for terms with positively charged fields}
\label{coefficient epsilon}
In the natural GUT, terms linear in a positively charged field play an important role in determining the VEVs of fields.
Therefore, if common suppression factor for terms with a positively charged field is introduced, the VEV relations in 
eq. (\ref{VEV}) do not change.
And terms which include two positively charged fields are important to determine the mass spectrum of superheavy 
particles. Therefore, if we introduce an independent suppression factor for terms with certain two positively charged fields
the gauge coupling unification conditions can be changed. In the following sub-sections, we consider this possibility.

%We focus on the renormalization group equation of gauge coupling. RGE depend on the mass spectrum of particles. Therefore, we discuss the coefficients of superpotential, which decide VEVs of negative charged fields.
Concretly we introduce the following suppression factors 
\begin{eqnarray}
&& W_{A^{\prime}}=\varepsilon_{A^{\prime}}\left(\lambda^{a^{\prime}+a} A^{\prime} A+\lambda^{a^{\prime}+3 a} A^{\prime} A^3\right), \\
&& W_{C^{\prime}}=\varepsilon_{C^{\prime}} \bar{C}\left(\lambda^{\bar{c}+c^{\prime}+a} A+\lambda^{\bar{c}+c^{\prime}+\bar{z}} \bar{Z}\right) C^{\prime}, \\
&& W_{\bar{C}^{\prime}}=\varepsilon_{\bar{C}^{\prime}} \bar{C}^{\prime}\left(\lambda^{\bar{c}^{\prime}+c+a} A+\lambda^{\bar{c}^{\prime}+c+z} Z\right) C, \\
&& W_{H^{\prime}}=\varepsilon_{H^{\prime}} \lambda^{h+h^{\prime}+a} H^{\prime} A H, \\
&& W_{X'Y'}=\varepsilon_{2 A^{\prime}} \lambda^{2 a^{\prime}} A^{\prime 2}+\varepsilon_{\bar{c}^{\prime} c^{\prime}} \lambda^{\bar{c}^{\prime}+c^{\prime}} \bar{C}^{\prime} C^{\prime}
+\varepsilon_{2 H^{\prime}} \lambda^{2 h^{\prime}} H^{\prime 2},
\end{eqnarray}
where the $F$-flatness conditions of the first four superpotentials determine the VEVs of negatively charged fields,
while the last superpotential are important to fix the mass spectrum of superheavy particles.
%The indices of $\varepsilon$ denote positive charged fields in each term. Here, we assign coefficients so that the VEVs of negative charged field do not depend on coefficients. If it is not so, there is a risk that $\langle A\rangle \gg \lambda^{-a}$ or $\langle A\rangle<\langle C\rangle$. In the former case, a term like $A^{\left(z_1+z_2+z_3\right) / a} Z_1 Z_2 Z_3$ becomes more dominant than terms with $\theta$. In the latter case, VEVs of negative charged fields are inconsistent with symmetry breaking scales. By using above superpotential, we obtain VEVs as
%\begin{eqnarray}
%&&\langle A\rangle  =v=\lambda^{-a}, \\
%&&\langle\bar{Z}\rangle  =\langle Z\rangle  =-\frac{3}{2} \lambda^{-z}, \\
%&&\langle C\rangle  =\langle\bar{C}\rangle=\lambda^{-\frac{1}{2}(c+\bar{c})},
%\end{eqnarray}
%We should also consider superpotentials which make mass matrices of heavy particles (\ref{mass5}),(\ref{mass10}) and (\ref{mass24}). 
The mass matrices of ${\bf 5}$ and ${\bf {\bar 5}}$ of $SU(5)$ become
\begin{equation}
M_I=\bordermatrix{
       &I_H & I_{H^\prime} & I_{C} &I_{C^\prime} \cr
\bar I_H & 0 & \varepsilon_{H^{\prime}}\lambda^{h+h^\prime}\alpha_I & 0 & 0 \cr
\bar I_{H^\prime} & \varepsilon_{H^{\prime}}\lambda^{h+h^\prime}\alpha_I & \varepsilon_{2H^{\prime}}\lambda^{2h^\prime} 
& 0 & \varepsilon_{H^\prime C^\prime}\lambda^{h'+c'+\frac{1}{2}(c-\bar c)} \cr
\bar I_{\bar C}& 0 & \varepsilon_{H^\prime}\lambda^{h'+\frac{3}{2}\bar c-\frac{1}{2}c} & 0 
& \varepsilon_{C^\prime}\lambda^{\bar c+c^\prime}\beta_I \cr
\bar I_{\bar C^\prime} &  \varepsilon_{{\bar C}^\prime}\lambda^{h+\bar c'-\frac{1}{2}(c-\bar c)} & \varepsilon_{H^\prime {\bar C}^\prime}\lambda^{h'+\bar c'-\frac{1}{2}(c-\bar c)}
 & \varepsilon_{\bar{C}^{\prime}}\lambda^{c+\bar c^\prime}\beta_I 
  & \varepsilon_{\bar{C}^{\prime} C^{\prime}}\lambda^{c^\prime+\bar c^\prime} \cr}.
\end{equation}
The determinants of reduced mass matrices, which are important to obtain the RGEs, 
 are written as %by $U(1)_A$ charges and coefficients of corresponding superpotentials as
\begin{eqnarray}
&& \operatorname{det} \bar{M}_{D^c} \sim \lambda^{2 h+2 h^{\prime}+c+\bar{c}+c^{\prime}+\bar{c}^{\prime}} \left[\varepsilon_{H^{\prime}}^2\varepsilon_{C^{\prime}} \varepsilon_{\bar{C}^{\prime}}\right] , \\
&& \operatorname{det} \bar{M}_L \sim \lambda^{2 h^{\prime}+c+\bar{c}+c^{\prime}+\bar{c}^{\prime}} 
\rm max\left[\varepsilon_{2 H^{\prime}}\varepsilon_{C^{\prime}} \varepsilon_{\bar{C}^{\prime}}, \,
 \varepsilon_{H^{\prime}}\varepsilon_{{\bar C}^{\prime}} \varepsilon_{H^\prime C^{\prime}},\,
  \varepsilon_{H^{\prime}}\varepsilon_{{\bar C}^{\prime}} \varepsilon_{H^\prime C^{\prime}}\lambda^{h-\frac{3}{2}c+\frac{1}{2}\bar{c}}\right].
\end{eqnarray}
%$[A \quad{\rm or}\quad B\quad{\rm or}\cdots]$ means to choose the largest one.

Similarly, the mass matrices of ${\bf 10}$ of $SU(5)$ become
\begin{equation}
M_I=\bordermatrix{
        &I_A & I_{A^\prime} &I_{ C}&I_{C^\prime} \cr
\bar I_A &0& \varepsilon_{A^{\prime}}\lambda^{a^\prime+a} \alpha_I & 0  & 
\varepsilon_{{C}^\prime}\lambda^{c^\prime-\frac{1}{2}(c-\bar c)+a} \cr
\bar I_{A^\prime} &\varepsilon_{A^{\prime}}\lambda^{a+a^\prime} \alpha_I & \varepsilon_{2A^{\prime}}\lambda^{2a^\prime} & 0 & 
\varepsilon_{A^\prime{C}^\prime}\lambda^{c^\prime-\frac{1}{2}(c-\bar c)+a^\prime} \cr
\bar I_{\bar C}&0 & 0 & 0 & \varepsilon_{C^{\prime}}\lambda^{\bar c+c^\prime}\beta_I \cr
\bar I_{\bar C^\prime} &\varepsilon_{{\bar C}^\prime}\lambda^{\bar c^\prime+\frac{1}{2}(c-\bar c)+a} &
\varepsilon_{A^\prime{\bar C}^\prime}\lambda^{\bar c^\prime+\frac{1}{2}(c-\bar c)+a^\prime}&
\varepsilon_{\bar{C}^{\prime}}\lambda^{c+\bar c^\prime}\beta_I & \varepsilon_{\bar{C}^{\prime} C^{\prime}}\lambda^{c^\prime+\bar c^\prime}\cr},
\end{equation}
and the determinants of reduced mass matrices are gives as
\begin{eqnarray}
&& \operatorname{det} \bar{M}_Q \sim \operatorname{det} \bar{M}_{U^c} \sim 
\lambda^{2 a^{\prime}+c+\bar{c}+c^{\prime}+\bar{c}^{\prime}} 
\left[\varepsilon_{2 A^{\prime}} \varepsilon_{C^{\prime}} \varepsilon_{\bar{C}^{\prime}}\right], \\
&& \operatorname{det} \bar{M}_{E^c} \sim \lambda^{2 a+2 a^{\prime}+c^{\prime}+\bar{c}^{\prime}} 
\rm max\left[\varepsilon_{A^{\prime}}^2 \varepsilon_{\bar{C}^{\prime} C^{\prime}},\,
\varepsilon_{2A^{\prime}} \varepsilon_{{C}^{\prime}}\varepsilon_{{\bar C}^\prime}\,
\varepsilon_{A^{\prime}} \varepsilon_{{\bar C}^{\prime}}\varepsilon_{A^\prime{C}^\prime},\,
\varepsilon_{A^{\prime}} \varepsilon_{{C}^{\prime}}\varepsilon_{A^\prime{\bar C}^\prime}\right].
\end{eqnarray}

For adjoint field $G,W,X$ and ${\bar X}$, mass matrices are given by
\begin{equation}
M_I=\bordermatrix{
            &    I_A       &        I_{A'}           \cr
\bar I_A    &     0        & \varepsilon_{A^{\prime}}\lambda^{a+a'}\alpha_I  \cr
\bar I_{A'} & \varepsilon_{A^{\prime}}\lambda^{a+a'}\alpha_I & \varepsilon_{2A^{\prime}}\lambda^{2a'}  \cr},
\end{equation}
and the determinants of the reduced mass matrices are
\begin{eqnarray}
&& \operatorname{det} \bar{M}_G \sim \operatorname{det} \bar{M}_W \sim \lambda^{2 a+2 a^{\prime}} \left[\varepsilon_{A^{\prime}}^2\right], \\
&& \operatorname{det} \bar{M}_X \sim \lambda^{2 a^{\prime}} \left[\varepsilon_{2 A^{\prime}}\right].
\end{eqnarray}
When we define the suppression parameters as 
\begin{eqnarray}
&&D_{D^c}=\left[\varepsilon_{H^{\prime}}^2\varepsilon_{C^{\prime}} \varepsilon_{\bar{C}^{\prime}}\right],\\
&&D_{L}={\rm max}\left[\varepsilon_{2 H^{\prime}}\varepsilon_{C^{\prime}} \varepsilon_{\bar{C}^{\prime}},\,
\varepsilon_{H^{\prime}}\varepsilon_{{\bar C}^{\prime}} \varepsilon_{H^\prime C^{\prime}},\,
\varepsilon_{H^{\prime}}\varepsilon_{{\bar C}^{\prime}} \varepsilon_{H^\prime C^{\prime}}\lambda^{h-\frac{3}{2}c+\frac{1}{2}\bar{c}}\right], \\
&&D_{Q,U^c}=\left[\varepsilon_{2 A^{\prime}} \varepsilon_{C^{\prime}} \varepsilon_{\bar{C}^{\prime}}\right],\\
&&D_{E^c}={\rm max}\left[\varepsilon_{A^{\prime}}^2 \varepsilon_{\bar{C}^{\prime} C^{\prime}},\,
\varepsilon_{2A^{\prime}} \varepsilon_{{C}^{\prime}}\varepsilon_{{\bar C}^\prime},\,
\varepsilon_{A^{\prime}} \varepsilon_{{\bar C}^{\prime}}\varepsilon_{A^\prime{C}^\prime},\,
\varepsilon_{A^{\prime}} \varepsilon_{{C}^{\prime}}\varepsilon_{A^\prime{\bar C}^\prime}\right],\\
&&D_{G,W}=\left[\varepsilon_{A^{\prime}}^2\right],\\
&&D_{X}=\left[\varepsilon_{2 A^{\prime}}\right],
\end{eqnarray}
the gauge coupling unification conditions can be rewritten as%, we can obtain the following two independent conditions
\begin{eqnarray}
&& \Lambda=\Lambda_{G} \left(\frac{D_{E^c}}{D_{Q,U^c}}\right)^{\frac{1}{6}}\left(\frac{D_{X}}{D_{G,W}}\right)^{\frac{1}{3}}\label{cutoff}
\label{Lambda}\\
&& \lambda^{2 h}=\left(\frac{D_{L}}{D_{D^c}}\right)\left(\frac{D_{Q,U^c}}{D_{E^c}}\right)^{\frac{2}{3}}\left(\frac{D_{G,W}}{D_{X}}\right)^{\frac{1}{3}}\label{2h}.
\label{h}
\end{eqnarray}
The heaviest neutrino mass can  be written as
\begin{eqnarray}
m_{\nu_{\tau}}&=&\lambda^{4+h+c-{\bar c}}\frac{\langle H_u\rangle^2}{\Lambda}\nonumber\\
&=&3\times10^{-6}\lambda^{c-{\bar c}}\left(\frac{D_{L}D_{Q,U^c}D_{G,W}}{D_{D^c}D_{E^c}D_{X}}\right)^{\frac{1}{2}}{\rm eV}.\label{nu_tau}
\label{TauNeutrino}
\end{eqnarray}

In the next subsection, using the above results, we discuss several possibilities to avoid the tension between 
the neutrino masses and the gauge coupling unification.

%%%%%%%%%%%%%%%%%%%%%%%%%%%%%%%%%%%%%%%%%%%%%%%%%%%%%%%%%%%%%%%%%%%%%%%%%%%%%%%%%%%%
\subsection{Models}
 In this subsection, we build several explicit natural GUT models which have no tension between the neutrino masses 
 and the gauge coupling unification by introducing various suppression factors as discussed in the previous subsection.
 %will investigate the $U(1)_A$ charge assignment, which is consistent with discussion of natural GUT. In the beginning, list the conditions that must be satisfied.
%\begin{eqnarray}
%2\psi_3+h=0 &:& \textrm{top Yukawa becomes $O(1)$},\label{Y_t}\\
%t-\psi_3-\frac{1}{2}(c-{\bar c})=\frac{5}{2} &:& \textrm{$Y_d$ to be correct size},\label{Y_d}\\
%\psi_3+t+c\geq0 &:& \textrm{mixing term of ${\bf {\bar 5}_{\bf \Psi_3}}$ and ${\bf {\bar 5}_T}$ is allowed},\label{mixing}\\
%\psi_3+\psi_1+2{\bar c}\geq0 &:& \textrm{right-handed neutrino mass matrix becomes rank3},\nonumber\\
%\label{right_nu}\\
%c+{\bar c}+a^\prime +a<0 &:& \textrm{${\bar C}A^\prime AC$ break DW form of $\langle A\rangle$},\label{CAAC}\\
%c+{\bar c}\geq0 &:& \textrm{Cabibbo angle to be correct size}.\label{Cabibbo}
%\end{eqnarray}
% Furthermore, we require $A^\prime A$ but exclude $A^\prime A^5$ so $-3a\leq a^\prime \leq-5a$. We choose $a=-1,a^\prime=3$, the least restrictive. Then the condition to realize DT splitting is $c+{\bar c}<-2$.
 
 First, let us explain the features common to the natural GUT models built in this paper. %we are about to propose. 
 They have no tension
 between the neutrino masses and the gauge coupling unification while they have all advantages
  of the usual natural GUTs except the basic principle that all terms allowed by the symmetry are introduced with
 $O(1)$ coefficients. We fix $a=-1$ and $a'=3$ which allow terms $A'A$ and $A'A^3$, and forbid $A'A^5$ and more to obtain the DW type VEV naturally,
% ofthe adjoint Higgs $A$, 
although we have another options to take $a=-1/2$ and $a=3/2$ which predict longer lifetime of 
 nucleon via dimension 6 operators because of larger unification scale. To obtain the realistic natural GUTs, they must
 satisfy the conditions listed in the end of the previous section, which can be rewritten as
 % From the last two conditions, we obtain
 \begin{eqnarray}
%2\psi_3+h=0 &:& \textrm{top Yukawa becomes $O(1)$},\label{Y_t}\\
&&t+\frac{1}{2}h-\frac{1}{2}(c-{\bar c})=\frac{5}{2}, \label{BiLarge}\\
&&-\frac{1}{2}h+t+c\geq0, \label{mixing}\\
&&-h+3+2{\bar c}\geq0,\label{RH}\\
 &&-6\leq c+{\bar c}<-2.\label{ccbar}
 \end{eqnarray} 
 And the three relations (\ref{Lambda})-(\ref{TauNeutrino}) we obtained in the end of the last subsection are 
 important to build the natural GUT models.

 %%%%%%%%%%%%%%%%%%%%%%%%%%%%%%%%%%%%%%%%%%
%\subsection{Model 1 : $\varepsilon_\nu\sim2.7\times10^{-4}$}
%In section (\ref{nu coefficient}), we showed that heaviest neutrino mass takes correct size at $h\sim-1$ by tuning of coefficient of right-handed neutrino mass term. Then $(\psi_1,\psi_2,\psi_3)=\left(\frac{7}{2},\frac{5}{2},\frac{1}{2}\right)$. Under this situation, necessary conditions to construct natural GUT are
% \begin{eqnarray}
% &&t-\frac{1}{2}-\frac{1}{2}(c-{\bar c})=\frac{5}{2},\\
% &&\frac{1}{2}+t+c\geq0,\\
% &&4+2{\bar c}\geq0,\\
% &&-6\leq c+{\bar c}<-2.
% \end{eqnarray}
 
%We can determine $U(1)_A$ charges assignments from these conditions as $(t,c,{\bar c})=\left(\frac{5}{2},-2,-1\right)$ or $\left(\frac{5}{2},-3,-2\right)$. For each case, we set $U(1)_A$ charge of other fields as Table\ref{assignment_nu1} and Table\ref{assignment_nu2}, consistent with natural GUT.

%%%%%%%%%%%%%%%%%%%%%%%%%%%%%%%%%%%%%%%%%%
\subsubsection{Model 2 : $\varepsilon_{H^\prime}\ll1, \textrm{others}\sim O(1)$}

In this model, we assume that $\varepsilon_{H^\prime}\ll1$ and the others are $O(1)$. 
The $\varepsilon_{H^\prime}$ dependence of determinants of the reduced mass matrices become
\begin{equation}
D_{D^c}\sim\varepsilon_{H^\prime}^2,\quad D_{L}\sim D_{Q,U^c}\sim D_{E^c}\sim D_{G,W}\sim D_{X}\sim1.
\end{equation}
%We can estimate cutoff $\Lambda$, $U(1)_A$ charge $h$, and heaviest neutrino mass $m_{\nu_\tau}$ by 
The eqs. (\ref{Lambda}),(\ref{h}) and (\ref{TauNeutrino}) are rewritten as 
\begin{eqnarray}
&&\Lambda\sim\Lambda_G,\\
&&\lambda^{2h}\sim\varepsilon_{H^\prime}^{-2},\\
&&m_{\nu_\tau}\sim3\times 10^{-6}\lambda^{c-\bar{c}}\varepsilon_{H^\prime}^{-1}{\rm eV}\sim 0.05 {\rm eV}.
\end{eqnarray}
From the last two relations in the above, we obtain
%To make $m_{\nu_\tau}$ consistent with experimental data, we should choose $c$ and ${\bar c}$ so that they satisfy $\lambda^{c-\bar{c}}\varepsilon^{-1}=\lambda^{-6}$. Therefore, we obtain a condition for $U(1)_A$ charges,
\begin{equation}
\label{h-c}
h=-(c-{\bar c})-6.
\end{equation}
Then, the condition (\ref{BiLarge}) becomes
\begin{equation}
t+h=-\frac{1}{2}.
\end{equation}
Among several solutions which satisfy all the conditions, two solution $(h,t,c,\bar c)=(-3,\frac{5}{2},-4,-1)$ and
$(h,t,c,\bar c)=(-3,\frac{5}{2},-3,0)$ have
the largest $h$ and an interesting feature that the half integer $U(1)_A$ charges play the same role as the R-parity.
An example of all $U(1)_A$ charges for the former solution is given in Table \ref{Model2}.
%If we choose $(h,t)=(-3,\frac{5}{2})$ as the solution, (\ref{mixing})-(\ref{Cabibbo}) and (\ref{h-c}) give the following condition for $c$ and ${\bar c}$.
%\begin{eqnarray}
%&&c\geq-4,\\
%&&{\bar c}\geq-3,\\
%&&-6\leq c+{\bar c}<-2,\\
%&&c-{\bar c}=-3.
%\end{eqnarray}
%We obtain solutions $(c,{\bar c})=(-4,-1)$ and $(-3,0)$, which satisfy the above conditions. As a result, we set $U(1)_A$ charges assignment as Table\ref{Model2}.
\begin{table}[t]
  \begin{center}
    \begin{tabular}{|c||c|c|c|}
      \hline
      $SO(10)$ & negatively charged fields & positively charged fields & matter fields \\ \hline \hline
      {\bf 45}  & $A(a=-1,-)$ & $A'(a'=3,-)$ & \\ \hline
      {\bf 16} & $C(c=-4,+)$ & $C'(c'=3,-)$  & $\Psi_i(\psi_1=\frac{9}{2}, \psi_2=\frac{7}{2}, \psi_3=\frac{3}{2}, +)$ \\ \hline
      ${\bf \overline{16}}$ & $\bar C(\bar c=-1,+)$ & $\bar C'(\bar c'=6,-)$ & \\ \hline
      {\bf 10} & $H(h=-3,+)$  & $H'(h'=4,-)$ & $T(t=\frac{5}{2},+)$ \\ \hline
      1 &  $Z(z=-2,-), \bar Z(\bar z=-2,-)$ & $S(s=5,+)$ &
      \\ \hline
    \end{tabular}
    \caption{$U(1)_A$ charge assignment in Model 2.}
    \label{Model2}
  \end{center}
\end{table}
%$U(1)_A$ charges of positive charged field are decided so that allowed superpotential terms are consistent with natural GUT. 
In these models, the suppression factor becomes
\begin{equation}
\varepsilon_{H^\prime}\sim\lambda^{-h}\sim\lambda^3\sim10^{-2}.
\end{equation}
The origin of this suppression factor can be understood by an approximate $Z_2$ symmmetry under which $H'$ is the unique
field with the odd $Z_2$ parity.

\begin{figure}[tbp]
\centering
\includegraphics[keepaspectratio,scale=0.4]{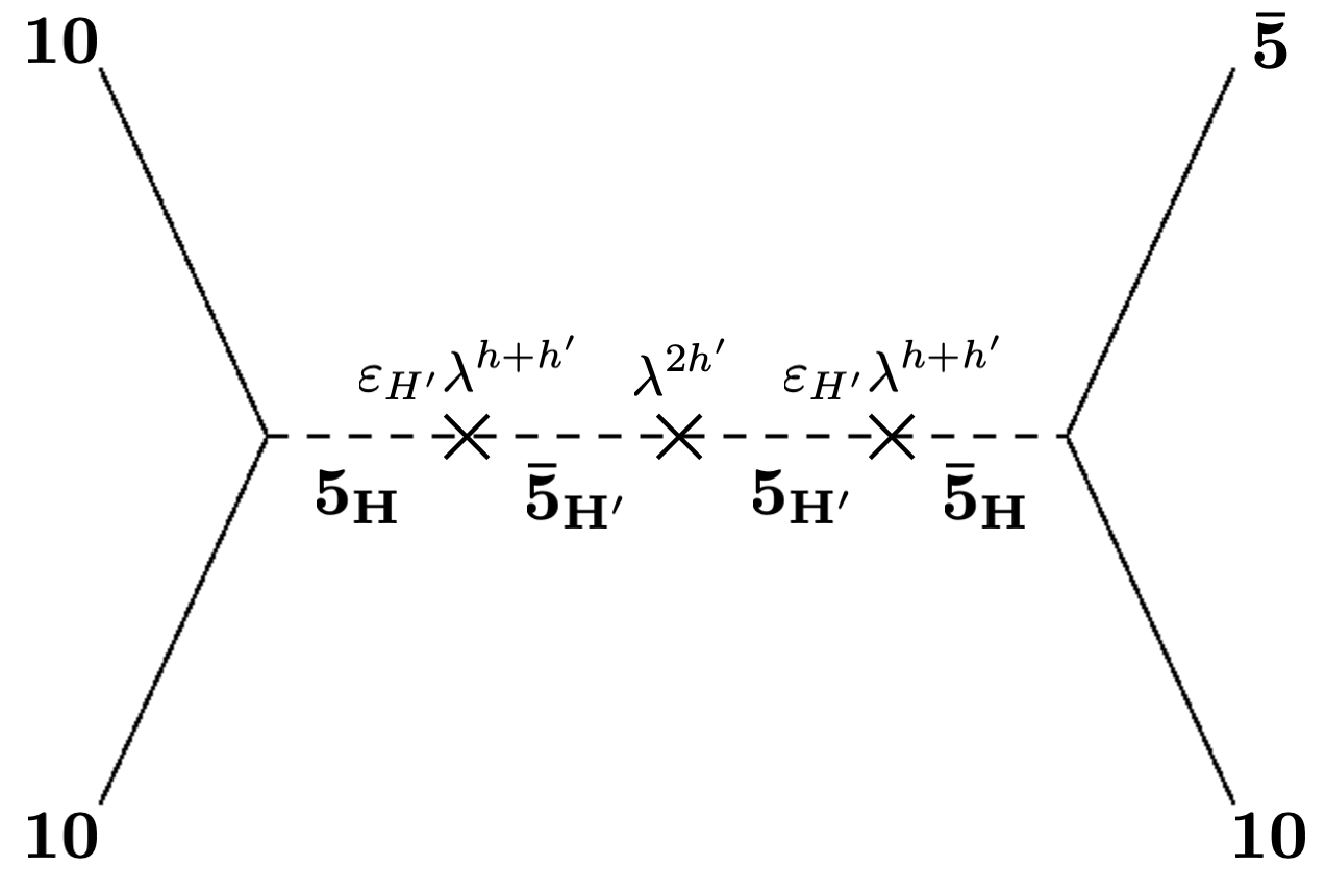}
\caption{proton decay mediated by colored Higgs}
\label{proton decay}
\end{figure}

%Since colored Higgs get masses of order $\varepsilon_{H^\prime}\lambda^{h+h^\prime}$, proton decay may be occur in this model. We consider proton decay through dim-5 operator (Figure\ref{proton decay}). 
Since the effective colored Higgs mass is calculated as
\begin{equation}
m_{H_T,eff}\sim\frac{\left(\varepsilon_{H^\prime}\lambda^{h+h^\prime}\right)^4}{\left(\varepsilon_{H^\prime}\lambda^{h+h^\prime}\right)^2\lambda^{2h^\prime}}\sim1,
\end{equation}
which equals to cutoff scale $\Lambda=\Lambda_G=2\times 10^{16}$ GeV, the signal for the nucleon decay via dimension 5
operators (see Fig. \ref{proton decay}) can be seen in future experiments although the predictions depend on the SUSY breaking scale and the
explicit structure of Yukawa couplings.
For example, the natural GUT with spontaneous SUSY breaking predicts quite large sfermion masses as
% SUSY breaking scale $m_{SUSY}$ and 
%typical sfermion mass $m_{¥tilde{f}}^2$ are
%\begin{eqnarray}
%&&m_{SUSY}\sim\frac{F_{\bar c}}{\Lambda}\sim10^{2-3}{\rm GeV},\\
%&&
$m_{\tilde{f}}^2\sim(10^{3-4}{\rm TeV})^2$ \cite{Maekawa:2019vzk}, 
%\end{eqnarray}
and therefore the proton decay via dimension 5 operators  is suppressed.

%%%%%%%%%%%%%%%%%%%%%%%%%%%%%%%%%%%%%%%%%%
\subsubsection{Model 3 : $\varepsilon_{H^\prime}\sim\lambda^\delta\varepsilon_{A^\prime}\ll1, \textrm{others}\sim O(1)$}
Here, we try to build a natural GUT in which not only the tension is avoided but also the cutoff scale becomes larger 
than $\Lambda_G$ since the cutoff scale is quite important in predicting the nucleon lifetime. 

In addition to $\varepsilon_{H^\prime}\ll1$, we introduce 
$\varepsilon_{A^\prime}\sim \lambda^{-\delta}\varepsilon_{H^\prime}\lesssim 1$ $(\delta>0)$ in order to change the cutoff.
%$\varepsilon_{H^\prime}:\varepsilon_{A^\prime}\sim1:\lambda^3$ and the others are $O(1)$. The $\varepsilon$ part of determinants of mass matrices are written as
Since
\begin{equation}
D_{D^c}\sim\varepsilon_{H^\prime}^2,\quad D_{G,W}\sim \varepsilon_{A^\prime}^2\sim\lambda^{-2\delta}\varepsilon_{H^\prime}^2,\quad D_{L}\sim D_{Q,U^c}\sim D_{E^c}\sim  D_{X}\sim1.
\end{equation}
the eqs. (\ref{Lambda}),(\ref{h}) and (\ref{TauNeutrino}) become
%The cutoff scale $\Lambda$, $U(1)_A$ charge $h$, and heaviest neutrino mass $m_{\nu_\tau}$ are estimated as
\begin{eqnarray}
&&\Lambda\sim\Lambda_G\lambda^{\frac{2}{3}\delta}\varepsilon_{H^\prime}^{-\frac{2}{3}},\label{Lambda3}\\
&&\lambda^{2h}\sim\lambda^{-\frac{2}{3}\delta}\varepsilon_{H^\prime}^{-\frac{4}{3}},\label{h3}\\
&&m_{\nu_\tau}\sim3\times 10^{-6}\lambda^{c-\bar{c}}\lambda^{-\delta}{\rm eV}\sim 0.05 {\rm eV}.\label{NuTau3}
\end{eqnarray}
Note that if $\lambda^\delta=\varepsilon_{H^\prime}$, the above conditions become nothing but those in the previous model.
Here, for simplicity, we assume that the cutoff is around the reduced Planck scale as 
$\Lambda\sim M_{Planck}\sim 2\times 10^{18}$ GeV. As a result, the first two equations (\ref{Lambda3}) and
(\ref{h3}) become 
\begin{eqnarray}
  &&\varepsilon_{H^\prime}\sim\lambda^{\delta+\frac{9}{2}}, \\
  &&h\sim -\delta-3.
\end{eqnarray}
The condition (\ref{NuTau3}) can be replaced as
\begin{equation}
  c-\bar c=\delta-6= -h-9.
\label{NuTau4}
\end{equation}
Substituting this relation into eq. (\ref{BiLarge}), we obtain
\begin{equation}
  t=-h-2.
\end{equation}
Among the solutions which satisfy the conditions (\ref{mixing})-(\ref{ccbar}), 
the solution with the largest $h$ has $(h=-4,t=2,c=-4,\bar c=1)$. Typical $U(1)_A$ charges can be seen
in Table \ref{Model3}.

\begin{table}[t]
  \begin{center}
    \begin{tabular}{|c||c|c|c|}
      \hline
      $SO(10)$ & negatively charged fields & positively charged fields & matter fields \\ \hline \hline
      {\bf 45}  & $A(a=-1,-)$ & $A'(a'=3,-)$ & \\ \hline
      {\bf 16} & $C(c=-4,+)$ & $C'(c'=1,-)$  & $\Psi_i(\psi_1=5, \psi_2=4, \psi_3=2, +)$ \\ \hline
      ${\bf \overline{16}}$ & $\bar C(\bar c=-1,+)$ & $\bar C'(\bar c'=6,-)$ & \\ \hline
      {\bf 10} & $H(h=-4,+)$  & $H'(h'=5,-)$ & $T(t=2,+)$ \\ \hline
      1 &  $Z(z=-2,-), \bar Z(\bar z=-2,-)$ & $S(s=3,+)$ &
      \\ \hline
    \end{tabular}
    \caption{$U(1)_A$ charge assignment in Model 3.}
    \label{Model3}
  \end{center}
\end{table}
The suppression factors are determined as
\begin{eqnarray}
&&\varepsilon_{H^\prime}\sim\lambda^{4.5+\delta}=\lambda^{5.5},\\
&&\varepsilon_{A^\prime}\sim\lambda^{-\delta}\varepsilon_{H^\prime}\sim\lambda^{4.5}.
%&&\Lambda\sim\Lambda_G\lambda^{-2}\sim4\times10^{17}{\rm GeV}.
\end{eqnarray}
Such suppression factors can be naturally realized by an approximate symmetries $Z_{2H'}$ and $Z_{2A'}$, 
where $Z_{2X}$ is a $Z_2$ symmetry under which only $X$ field has odd parity. When these approximate symmetries are
imposed, the other suppression factors appear, for example, $\varepsilon_{A^\prime\bar C'}$, but these additional 
suppression factors do not change the physical results.

The effective colored Higgs mass related proton decay via dim-5 operators becomes
\begin{equation}
m_{H_T,eff}\sim\Lambda\varepsilon^2\lambda^{2h}\sim\Lambda_G\sim2\times10^{16}{\rm GeV},
\end{equation}
which means that the proton decay via dimesnion 5 operators can be seen in future experiment although the predictions
of the lifetime depends on the Yukawa structure and the SUSY breaking scale.

It is a general result that the effective colored Higgs mass become around $\Lambda_G$. Actually, it is shown only by
eqs.(\ref{Lambda3})-(\ref{NuTau3}). In the next subsubsection, we try to build the natural GUT with larger effective
colored Higgs mass.

%More generally, even assuming $\varepsilon\equiv\varepsilon_{H^\prime}\ll1$, $\varepsilon_{H^\prime}:\varepsilon_{A^\prime}\sim1:\lambda^\delta$ and the others are $O(1)$, effective colored Higgs mass remains $\lambda_G$. In that case, $\Lambda\sim\Lambda_G\lambda^{\frac{2\delta}{3}}\varepsilon^{-\frac{2}{3}}, \lambda^{2h}\sim\lambda^{-\frac{2\delta}{3}}\varepsilon^{-\frac{4}{3}}$ and we obtain effective colored Higgs mass $m_{H_T,eff}\sim\Lambda\varepsilon^2\lambda^{2h}\sim\Lambda_G$. Therefore, $\varepsilon_{H^\prime}$ and $\varepsilon_{A^\prime}$ alone cannot make $m_{H_T}$ large.

%%%%%%%%%%%%%%%%%%%%%%%%%%%%%%%%%%%%%%%%%%%%%%%%%%%%%%%%%%%%%%%%%%%%%%%%%%%%%%%%%%%%%%
\subsubsection{Model 4 : $\varepsilon_{2H^\prime}\ll\varepsilon_{H^\prime}\ll1, \textrm{others}\sim O(1)$}

Generically, the effective colored Higgs mass can be obtained as
\begin{equation}
  m_{H_T,eff}\sim\frac{\left(\varepsilon_{H^\prime}\lambda^{h+h^\prime}\right)^4}
  {\left(\varepsilon_{H^\prime}\lambda^{h+h^\prime}\right)^2\varepsilon_{2H^\prime}\lambda^{2h^\prime}}\Lambda
  \sim\varepsilon_{H^\prime}^2\varepsilon_{2H^\prime}^{-1}\lambda^{2h}\Lambda
  \sim\varepsilon_{H^\prime}^2\varepsilon_{2H^\prime}^{-1}\left(\frac{D_L}{D_{D^c}}\right)\left(\frac{D_{Q,U^c}}{D_{E^c}}\right)^{\frac{1}{2}}\Lambda_G,
\end{equation}
where the last similarity is shown by eqs.(\ref{Lambda}) and (\ref{h}). Obviously, $\frac{D_L}{D_{D^c}}\geq\frac{\varepsilon_{2H^\prime}} {\varepsilon_{H^\prime}^2}$ and 
$\frac{D_{Q,U^c}}{D_{E^c}}\leq 1$. Therefore, to obtain the 
larger effective colored Higgs mass than $\Lambda_G$, it is sufficient that $\frac{D_L}{D_{D^c}}>\frac{\varepsilon_{2H^\prime}} {\varepsilon_{H^\prime}^2}$ and
$\frac{D_{Q,U^c}}{D_{E^c}}\sim 1$. This is possible if
$\varepsilon_{2H^\prime}\ll\varepsilon_{H^\prime}\ll1, \textrm{others}\sim O(1)$.
% because 
%$D_{L}={\rm max}\left[\varepsilon_{2 H^{\prime}}\varepsilon_{C^{\prime}} \varepsilon_{\bar{C}^{\prime}},\,
%\varepsilon_{H^{\prime}}\varepsilon_{{\bar C}^{\prime}} \varepsilon_{H^\prime C^{\prime}},\,
%\varepsilon_{H^{\prime}}\varepsilon_{{\bar C}^{\prime}} \varepsilon_{H^\prime C^{\prime}}
%\lambda^{h-\frac{3}{2}c+\frac{1}{2}\bar{c}}\right]\sim \varepsilon_{H^{\prime}}\gg\varepsilon_{2 H^{\prime}}$.
%We construct the model in which colored Higgs is heavy so that proton decay can be suppressed. In this assumption, the $\varepsilon$ part of determinants of mass matrices are written as
Since 
\begin{equation}
D_{D^c}\sim\varepsilon_{H^\prime}^2,\quad D_{L}\sim\varepsilon_{H^\prime},\quad D_{Q,U^c}\sim D_{E^c}\sim D_{G,W}\sim D_{X}\sim1,
\end{equation}
we obtain the effective colored Higgs mass as
%can estimate $\Lambda\sim\Lambda_G$ and $\lambda^{2h}\sim\varepsilon_{H^\prime}^{-1}$, so effective colored Higgs mass is
\begin{equation}
m_{H_T,eff}
%\sim\frac{\left(\varepsilon_{H^\prime}\lambda^{h+h^\prime}\right)^4}{\left(\varepsilon_{H^\prime}\lambda^{h+h^\prime}\right)^2\varepsilon_{2H^\prime}\lambda^{2h^\prime}}\Lambda
\sim\frac{\varepsilon_{H^\prime}}{\varepsilon_{2H^\prime}}\Lambda_G>\Lambda_G.
\end{equation}
The explicit $U(1)_A$ charge assignment can be shown in Table \ref{Model2}, which is the same as in the model 2.
From the eqs. (\ref{Lambda}) and (\ref{h}), we can obtain
%In this model, the condition to $U(1)_A$ charge from $m_{\nu_\tau}$ is written
\begin{eqnarray}
  &&\Lambda=\Lambda_G, \\
  &&\varepsilon_{H'}=\lambda^{-2h}=10^{-4}.
%  c+{\bar c}+h=-6,
\end{eqnarray}
Since the unification scale $\lambda^{-a}\Lambda<\Lambda_G$, the nucloen decay via dimension 6 operators may be seen 
in future experiments, while the proton decay via dimension 5 operators is suppressed although it strongly depends on
$\varepsilon_{2H'}$.
%it is same with Model 2. Therefore, we can assign $U(1)_A$ charge exactly same
%with Model 2.

%As a result, we set coefficient as $\varepsilon_{H^\prime}\sim\lambda^6\sim10^{-4}$ and obtain effective colored Higgs mass
%\begin{equation}
%m_{H_T,eff}\sim\frac{2\times10^{12}}{\varepsilon_{2H^\prime}}{\rm GeV}.
%\end{equation}
%We do not have restriction for $\varepsilon_{2H^\prime}$.

Unfortunately, these suppression factors cannot be realized by an approximate symmetry.
For example, an approximate symmetry, in which only $H'$ has non-trivial charge, results in 
$\varepsilon_{{\bar C}'H'}\sim \varepsilon_{H'}$ and $\varepsilon_{H'}^2\lesssim \varepsilon_{2H'}$,
and therefore, we obtain 
$\frac{D_L}{D_{D^c}}\sim\frac{\varepsilon_{2H^\prime}} {\varepsilon_{H^\prime}^2}$.

%%%%%%%%%%%%%%%%%%%%%%%%%%%%%%%%%%%%%%%%%%%%%%%%%%%%%%%%%%%%%%%%%%%%%%%%%%%%%%%%%%%%%%
\section{Discussion and summary}
Under the natural assumption that all terms allowed by the symmetry are introduced by $O(1)$ coefficients, the natural GUT solves various problem of SUSY GUT and gives a GUT that leads to the Standard Model, which is consistent with almost all observations and experiments. 
Unfortunately, the natural GUT has an unsatisfied point that many $O(1)$ coefficients must be artificially chosen between 0.5 and 2 to achieve unification of the gauge coupling constants. This problem is due to the fact that the neutrino masses become too small to satisfy the measured values under the conditions for unification of the gauge coupling constants.

In this paper we discussed how to avoid the tension between the unification of gauge coupling constants and neutrino masses in the natural GUT. 
In particular, we considered the possibilities that the tension could be eliminated by assuming that, for some reason, some terms have suppression factors in addition to the suppression factors determined by the $U(1)_A$ symmetry. 
We found several solutions and explicitly built natural GUT models .
%corresponding to them. 
For some solutions, we also found that their additional suppression factors can be understood naturally with approximate symmetries. 

%We focused on how nucleon decay, an important prediction of GUT, changes in these solutions. We found that in most of solutions the nucleon decay due to operators of dimension 5, which is quite suppressed in the original natural GUT,  is in the range that can be reached by future experiments, although we also showed that the nucleon decay can be suppressed in a natural GUT with suppression factors which cannot be understood by an approximate symmetry. Moreover, it was suggested that proton decay due to dim. 6 operators generally occurs more often in the original natural GUT than in the usual GUT and can be seen experimentally in the near future, but it was shown that it does not necessarily occur more often in the natural GUT with suppression factors.

We focused on how nucleon decay, an important prediction of GUT, changes in these solutions. In the original SUSY GUT scenario, the nucleon decay via dim. 5 operators is
important while the nucleon decay via dim. 6 operators is suppressed because of the larger unification scale. In the original natural GUT scenario, the nucleon decay via dim. 6 operators becomes interesting because the unification scale
becomes generally lower, while the nucleon decay via dim. 5 operators is strongly suppressed because the effective colored Higgs mass becomes $\lambda^{2h}\Lambda_G$ with $h=-3$. This is an important prediction in the natural GUT.  In the natural GUT with suppression factors, which is discussed 
in this paper to avoid the tension between the gauge coupling unification and the neutrino
masses, these predictions for nucleon decay can change. The model with suppression factor for terms for right-handed neutrino masses gives similar predictions on nucleon decay as the original natural GUT because the colored Higgs mass becomes $\lambda^{2h}\Lambda_G$ with $h=-1$.
%off scale is around the usual GUT scale $\Lambda_G$ and the $U(1)_A$ charge of Higgs is still negative.  
In the models with suppression factors explained by a symmetry for terms with positively charged fields, the nucleon decay via dim. 5 operators becomes more important generically, while
the nucleon decay via dim. 6 operators can be suppressed. This is an important observation
in this paper, although we also showed that the nucleon decay via dim. 5 operators can be suppressed in a natural GUT with suppression factors which cannot be understood by an approximate symmetry.

Note that in the natural GUT, in which the suppression factors can be understood by an approximate symmetry, the suppression factor discussed above cannot be understood in terms of spontaneous breaking of the symmetry.
%, e.g., the $Z_2$ symmetry.
For example, if we try to explain the suppression factor of the $H'H$ term, where $H'$ and $H$ have odd and even $Z_2$ parity, respectively, by the VEV of the $Z_2$ odd and $U(1)_A$ negatively charged field $Z_-$ from a symmetric term $\lambda^{z_-+h'+h}Z_-H'H$, the suppression factor does not appear
because the VEV relation $\VEV{Z_-}\sim \lambda^{-z_-}$ is cancelled by the enhanced factor $\lambda^{z_-}$. Therefore, the approximate symmetries which may appear in the natural GUT with suppression factors must be understood by other reason,
for example, extra-dimension, additional $U(1)_A'$ symmetry, or other stringy reason,
etc. 
%Even if the suppression factor cannot be understood by an approximate symmetry, 
%these other reasons may realize the suppression factors. 
We hope that our consideration can be a hint to find the model beyond the natural GUT. 

\section{Acknowledgement}
N.M. thanks K. Chahara and T. Himekawa for the collaborations and discussions in the early stage of this work.
This work is supported in part
by the Grant-in-Aid for Scientific Research from the
Ministry of Education, Culture, Sports, Science and
Technology in Japan No. 19K03823 (N.M.).

\end{document}